\documentclass[pdflatex,sn-nature]{sn-jnl}% Basic Springer Nature Reference Style/Chemistry Reference Style
\usepackage{graphicx}%
\usepackage{multirow}%
\usepackage{amsmath,amssymb,amsfonts}%
\usepackage{amsthm}%
\usepackage{mathrsfs}%
\usepackage[title]{appendix}%
\usepackage[table]{xcolor}% must come before tabularx
\usepackage{textcomp}%
\usepackage{manyfoot}%
\usepackage{booktabs}%
\usepackage{algorithm}%
\usepackage{algorithmicx}%
\usepackage{algpseudocode}%
\usepackage{listings}%
\usepackage{array}% recommended for tabularx
\usepackage{tabularx}%

\theoremstyle{thmstyleone}%
\theoremstyle{thmstyletwo}%

\theoremstyle{thmstylethree}%

\begin{document}

\title[Article Title]{Physiologically Informed Digital Auscultation for Pneumonia Detection in Long-term Care Residents}


\author*[1]{\fnm{Nicholas} \sur{Rasmussen}}\email{nrasmus@uw.edu}

\author[1]{\fnm{Oleg} \sur{Zaslavsky}}\email{ozasl@uw.edu}

\author[2]{\fnm{Zih-Ling} \sur{Wang}}\email{zihlingw@uw.edu}

\author[2]{\fnm{Hongyu} \sur{Yu}}\email{hy227@uw.edu}

\author[1]{\fnm{Joelle} \sur{Fathi}}\email{thirsk@uw.edu}

\author[3]{\fnm{Kaibao} \sur{Nie}}\email{niek@uw.edu}

\author[4]{\fnm{Amil} \sur{Khanzada}}\email{amil@virufy.org}

\author[5]{\fnm{Tomoko} \sur{Ito}}\email{tito@md.tsukuba.ac.jp}

\affil[1]{\orgdiv{Biobehavioral Nursing \& Health Informatics, School of Nursing}, \orgname{University of Washington}, \orgaddress{\street{1959 NE Pacific St}, \city{Seattle}, \state{WA}, \country{USA}}}

\affil[2]{\orgdiv{School of Nursing}, \orgname{University of Washington}, \orgaddress{\street{1959 NE Pacific St}, \city{Seattle}, \state{WA}, \country{USA}}}

\affil[3]{\orgdiv{Electrical Engineering}, \orgname{University of Washington Bothell}, \orgaddress{\street{17827 113$^{th}$ Ave NE}, \city{Bothell}, \state{WA}, \country{USA}}}

\affil[4]{ \orgname{Virufy (The Covid Detection Foundation)}, \orgaddress{\city{Los Altos}, \state{CA}, \country{USA}}}

\affil[5]{\orgdiv{Institute of Medicine}, \orgname{University of Tsukuba}, \orgaddress{\street{1-1-1 Tenno-dai}, \city{Tsukuba}, \state{Ibaraki}, \country{Japan}}}


\abstract{
Pneumonia is difficult to diagnose in older long-term care residents; multimorbidity and atypical presentations obscure signs, motivating operationally efficient objective testing. We analyzed multi‑channel digital stethoscope recordings from 185 Japanese residents (73 pneumonia, 112 symptomatic without), using radiologist‑confirmed chest X‑rays and clinician diagnoses as supervisory signals that train convolutional neural networks, multimodal fusion, and channel‑based variants with time-domain Grad‑CAM interpretability. Models were evaluated with repeated patient-level cross-validation showing models with X‑ray supervision outperformed clinician supervision (F1 0.729, accuracy 0.783 vs. F1 0.637, accuracy 0.711). Additionally, a three‑channel selection protocol maintained performance (F1 0.736; accuracy 0.803), with two mid‑thoracic sites ranking highest and Grad‑CAM attention overlapping adventitious sounds. These findings indicate automated multi‑channel lung‑sound analysis can aid long-term care pneumonia diagnosis, with X‑ray supervision being more reliable than clinical, and fewer channels preserving performance while lowering acquisition times.
}

\keywords{Pneumonia Detection, Biomedical Signal Processing, Channel Selection, Skilled Nursing Facilities, Geriatric Care, Machine Learning}



\maketitle

\section{Introduction}

Pneumonia is a major cause of morbidity and mortality in long‑term care (LTC), with incidence and short‑term mortality far exceeding community rates. Frailty, multimorbidity, cognitive impairment, and communal living markedly increase susceptibility and severity \cite{Mylotte2020JAMDA, Cao2022Cochrane, hoogendijk2019frailty}. Presentations are often non-classic, with confusion, lethargy, or functional decline without fever or cough helping drive missed or delayed diagnosis \cite{Cilloniz2026Chest, Aliberti2021Lancet, vaughn2024community}. Limited on‑site testing equipment, constrained trained workforce capacity, difficulty of transporting frail residents, and radiographic limitations—including positioning, portable image quality, and interpretive variability—further heighten diagnostic uncertainty \cite{Spector2013MedCare, Estrada2024JAMAOpen}. Thus, practical bedside diagnostics deployable within LTC facilities are urgently needed to improve outcomes \cite{Mylotte2020JAMDA, veronese2025infectious}. 

Lung auscultation is central to bedside respiratory assessment, but human interpretation of adventitious sounds shows limited sensitivity and variable documentation even in hospitalized pneumonia \cite{Arts2020SciRep, vaughn2024community}. Reliability is further constrained by variation in clinical expression, ambient noise, and the transient nature of pathological sounds \cite{heitmann2023deepbreath, kim2022coming, kim2021respiratory}. These limitations are amplified in LTC, where initial assessments are often performed by nursing staff without prescriber oversight, contributing to diagnostic uncertainty and inappropriate antibiotic use \cite{Mylotte2020JAMDA,jump2018infectious}.

Digital stethoscopes and automated lung‑sound analysis standardize acquisition and enable machine‑learning models to extract clinically relevant acoustic signatures \cite{heitmann2023deepbreath,garcia2023machine}. Early studies demonstrate promising performance across respiratory conditions and can reduce inter‑observer variability while ensuring reproducible, auditable outputs \cite{Kapetanidis2024Sensors,Landry2025ERR}. However, much of the literature relies on public repositories or case‑control designs, and reported near‑perfect accuracies often reflect dataset artifacts such as correlations with recording site, equipment, or protocol \cite{Roberts2017CV}.

\begin{figure}[t!]
\centering
\includegraphics[width=1.00\linewidth]{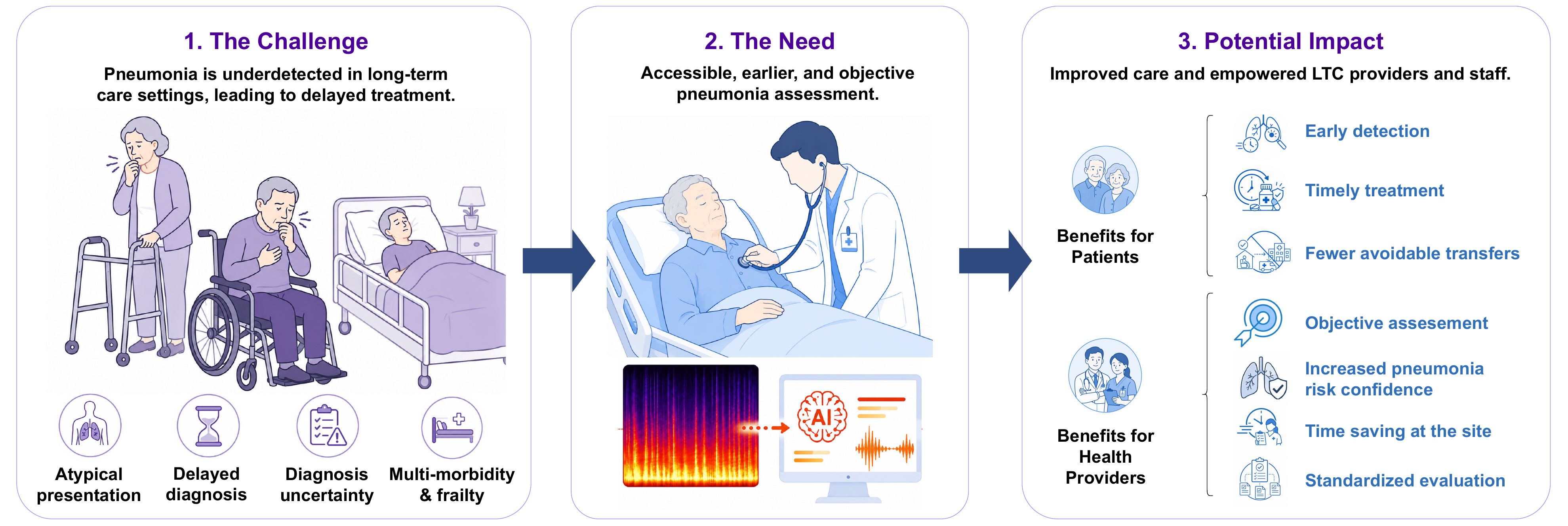}
\caption{Overview of the clinical challenge, need, and potential impact of AI‑assisted digital auscultation for pneumonia detection in LTC settings. The infographic illustrates how under‑recognition of pneumonia in frail older adults motivates the development of accessible, multi‑channel digital stethoscope assessment, enabling earlier diagnosis, timely treatment, and standardized evaluation for both patients and providers.}

\label{fig:intro}
\end{figure}

Prospective elderly adult evidence is limited and largely absent for LTC residents. Prior cohorts such as pediatric, emergency‑department, or mobile‑phone recordings do not reflect nursing‑home conditions, where high comorbidity, cognitive impairment, and frequent immobility shape both acoustic expression and protocol feasibility \cite{heitmann2023deepbreath, Huecker2026BMCPulm, Mylotte2020JAMDA}. These factors constrain the applicability of multi‑site recording protocols used elsewhere in the literature and motivate targeted prospective cohorts and protocols.

Practical deployment in LTC requires two advances. First, models must be trained and validated on patient‑level cohorts drawn from the target population to ensure performance generalizes to the intended clinical environment. Second, recording protocols must balance diagnostic yield with bedside feasibility: multi‑site auscultation can capture spatially localized pathology but is often impractical for frail or bedridden residents, increases staff workload, and may require uncomfortable or unsafe repositioning \cite{Tandan2020JAMDA}. Thus, a reduced‑channel protocol that preserves diagnostic signal while minimizing burden would improve real‑world adoption \cite{kim2025enhanced}.

In this study, we evaluate whether standardized multi‑channel digital lung sounds can detect pneumonia in older LTC residents. Our primary objective was to assess machine‑learning performance for distinguishing pneumonia from symptomatic non‑pneumonia using patient‑level partitions under two supervisory signals: radiologist‑confirmed chest‑X‑ray evidence and clinician‑diagnosed pneumonia. Next, we examine the contribution of individual auscultation sites and test whether a reduced‑channel protocol could preserve diagnostic performance while lowering bedside burden. Finally, we assessed whether model attention aligned with respiratory phases and annotated adventitious sounds ensuring an interpretable detection model. Together, these analyses assess the diagnostic potential and practical feasibility of automated lung‑sound analysis in LTC.

\begin{figure}[t!]
\centering
\includegraphics[width=.99\linewidth]{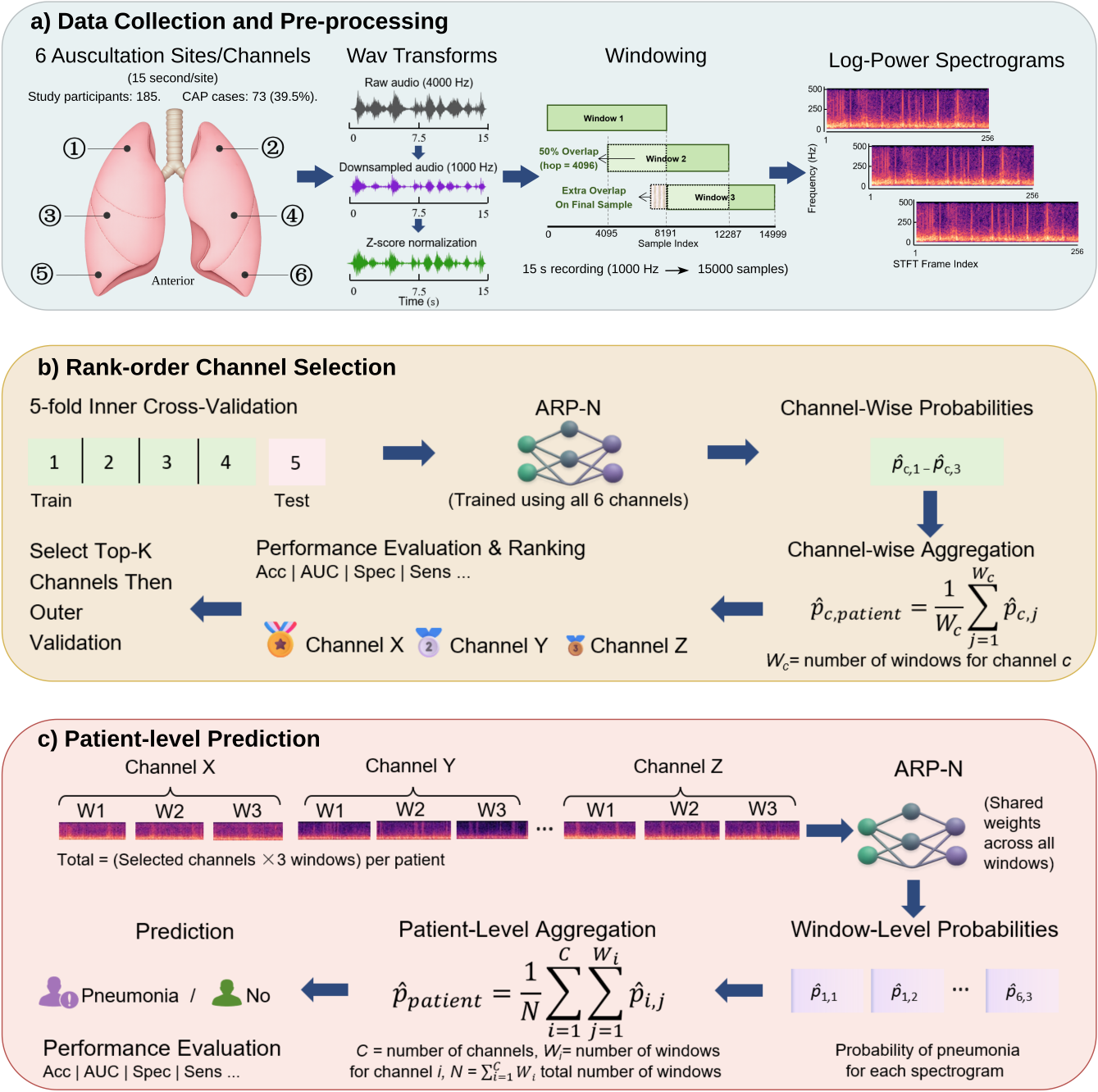}
\caption{ Workflow for multi-channel auscultation analysis.
\textbf{(a)} Study cohort and six anterior auscultation sites with their downsampling, normalization, window segmentation, and STFT representation.
\textbf{(b)} Rank-order Channel Selection via 5-fold inner cross-validation.
\textbf{(c)} Prediction using the shared-weight ARP-N CNN, with window probabilities aggregated to a patient-level score.
}
\label{fig:mp}
\end{figure}

\section{Results}

\subsection{Cohort Analysis}

\paragraph{Cohort Composition.} Among 185 geriatric patients, 112 (61\%) were classified as non‑pneumonia and 73 (39\%) as pneumonia by the reviewing clinician, informing the cohort stratification reported in Table~\ref{Table:Cohort}, while radiographic opacities were present in 77 patients. Radiographic findings contributed to, but did not determine, the clinical diagnosis, and disagreement between the two labels was expected.

\paragraph{Physiologic Measures.} Relative to non-pneumonia patients, patients with pneumonia showed modest reductions in oxygen saturation (94.7~$\pm$~3.67\% vs.\ 95.7~$\pm$~2.76\%; OR~0.82--1.00, $p=0.05$) and higher pulse rates (81.7~$\pm$~15.5 vs.\ 76.6~$\pm$~13.7\,bpm; OR~1.00--1.04, $p=0.02$) consistent with clinical expectations. Visual analogue scale (VAS) scores demonstrated the strongest between-group separation (34.8~$\pm$~26.7 vs.\ 22.8~$\pm$~18.4; OR~1.01--1.03, $p=0.001$). Other physiologic measures showed only modest group differences.

\paragraph{Categorical Variables.} Categorical variables also differed between groups. White sputum (OR~1.61--7.25, $p=0.001$), thick sputum (OR~1.71--6.64, $p=0.001$), and constant dyspnea (OR~1.12--7.95, $p=0.028$) were more common among pneumonia patients. Comorbidities---including asthma, COPD, cancer, heart failure, dementia, and interstitial pneumonia---showed no strong univariate associations, with sparse category counts producing wide confidence intervals (e.g., 0--$\infty$). Tracheostomy showed a borderline association (OR~0.018--1.13, $p=0.065$), most likely attributable low prevalence.

\subsection{Primary Diagnostic Performance}

\paragraph{XRAY vs.\ Diag Labels.} As shown in Table~\ref{tab:MR}, XRAY supervision produced stronger and more stable models than Diag labels across all architectures, with the largest gains observed in stethoscope-based learning. Under XRAY labels, the stethoscope-only model achieved the highest F1 score (0.729), accuracy (0.783), negative predictive value (NPV; 0.789), sensitivity (0.686), and positive predictive value (PPV; 0.805), suggesting that acoustic features carry discriminatory information for the pneumonia classification task. These findings are consistent with the noisier and more heterogeneous nature of subjective clinician-assigned diagnostic labels.

\paragraph{Multimodal Fusion.} The multimodal clinical+stethoscope model exhibited a complementary performance pattern. Although the stethoscope‑only model led most XRAY metrics, the fusion model achieved the highest AUC (.791) and specificity (.867), indicating that structured clinical variables can sharpen decision boundaries at particular operating points. This was most evident in the threshold‑based sensitivity metrics. Specificity at 90\% Sensitivity (S90) and Specificity at 80\% Sensitivity (S80) both improved under Diag and XRAY supervision.

\paragraph{Variability Patterns.} Variability analyses further supported XRAY supervision. The stethoscope-only model exhibited the lowest intra-split variability for F1 (SD = 0.067), accuracy (SD = 0.071), NPV (SD = 0.050), and sensitivity (SD = 0.076), indicating stable behavior across patient-level partitions. The fusion model showed similarly low inter-split variability for AUC (SD = 0.016), specificity (SD = 0.017), and sensitivity (SD = 0.019), suggesting that the addition of clinical features helped regularize decision thresholds.

\begin{figure}[h!]
\centering
\includegraphics[width=.95\linewidth]{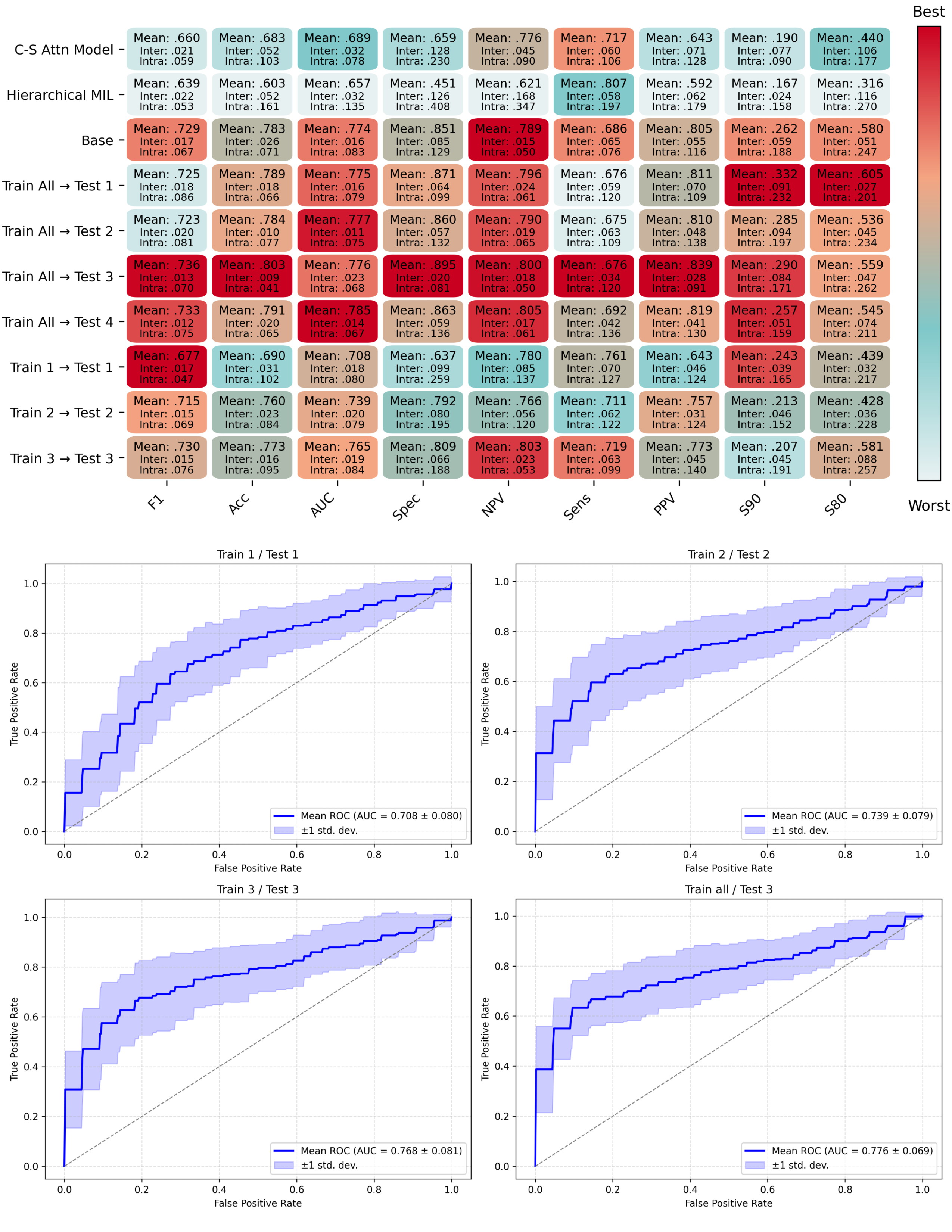}
\caption{
\textbf{Top: Channel-selection ablation heatmap.} Heatmap values show per-metric signal-to-noise ratio (SNR), computed as \( \mathrm{SNR} = \frac{\mathrm{mean}}{\sqrt{2\,\mathrm{inter\text{-}fold} + \mathrm{intra\text{-}fold}}}. \) Cells report cross-validated mean and inter-/intra-fold standard deviations, normalized per column to enable direct comparison of channel robustness. A unified teal--coral colormap encodes relative stability of performance; warmer tones indicate greater discriminative consistency. \textbf{Bottom: ROC curves across training configurations.} Mean ROC with $\pm 1$\,SD bands (25 folds) shows improved discrimination and reduced variance as additional stethoscope channels are added to training.
}

\label{fig:abb}
\end{figure}

\subsection{Channel Ablations}

\paragraph{Architecture Comparison.} Figure~\ref{fig:abb} shows that across the three channel-aware architectures, the Mean-Probability ARP--N (Base) model was the strongest and most stable (Fig.~3). It achieved the highest overall discrimination (F1 = 0.729, accuracy = 0.783, AUC = 0.774, specificity = 0.851) and consistently low across-fold variability, indicating a robust representation under patient-level splits. By contrast, architectures that imposed cross-channel spatial structure (Channel-Stacked Attention, C-S Attn Model) or treated channels sequentially (Hierarchical multiple-instance learning, Hierarchical MIL) showed markedly weaker discrimination (AUC = 0.689 and 0.657, respectively) and substantially lower specificity, with higher intra-split variability reflecting unstable operating points.

\paragraph{Train-all, Test-subset Ablations.} Channel ablations indicated that the pneumonia signal was concentrated in a small subset of auscultation sites. Training on all six channels while evaluating on restricted subsets yielded the strongest operating-point performance for three- and four-channel inference (Train All $\rightarrow$ Test~3/4), with F1 = 0.736/0.733, accuracy = 0.803/0.791, and specificity = 0.895/0.863. These configurations also showed the lowest inter-split variability (e.g., Test~3 accuracy SD = 0.009, F1 SD = 0.013). Even single-channel inference (Train All $\rightarrow$ Test~1) preserved high specificity at clinically relevant sensitivity thresholds (S90 = 0.332; S80 = 0.605).

\paragraph{Matched Train/Test Ablations.} When models were trained and evaluated on the same restricted channel sets, performance shifted predictably. Single-channel models (Train~1 $\rightarrow$ Test~1) produced high sensitivity (0.761) and strong NPV (0.780) but lower specificity (0.637) and AUC (0.708), reflecting recall-heavy behavior driven by limited spatial coverage. Two-channel training improved the balance across metrics (F1 = 0.715, accuracy = 0.760, AUC = 0.739), and three-channel training (Train~3 $\rightarrow$ Test~3) recovered much of the full-model discrimination (F1 = 0.730, accuracy = 0.773, AUC = 0.765) while delivering strong specificity (0.809) with moderate variability.

\subsection{XAI: Channel Importance and Attention Alignment}

\paragraph{Channel Importance.} Across 25 independent rankings, Channels~3 and~4 formed a dominant tier. Channel~3 was the modal top-ranked site and achieved the highest Borda score (119), followed by Channel~4 (101). Channel~6 (88) occupied a stable mid-high tier, Channels~2 and~1 ranked variably across seeds (79 and 71), and Channel~5 was consistently lowest (67). This hierarchy was robust across seeds, folds, and label sets, indicating repeated convergence on the same physiologically plausible auscultation placement when sampled at 1~kHz.

\paragraph{Attention Alignment with Expert Annotations.} Spectrogram-based attention consistently concentrated on the acoustic regions marked by all four annotators. On average, attention covered 74\% of inspiratory events, 64\% of expiratory events, 61\% of rales, 82\% of rhonchi, and 81\% of wheezes across Annotators~A--D. Alignment was strongest for the pulmonary nurse practitioner  (Annotator~D: inspirations 82\%, expirations 69\%, rales 71\%, rhonchi 81\%).

\begin{figure}[t!]
\centering
\includegraphics[width=.97\linewidth]{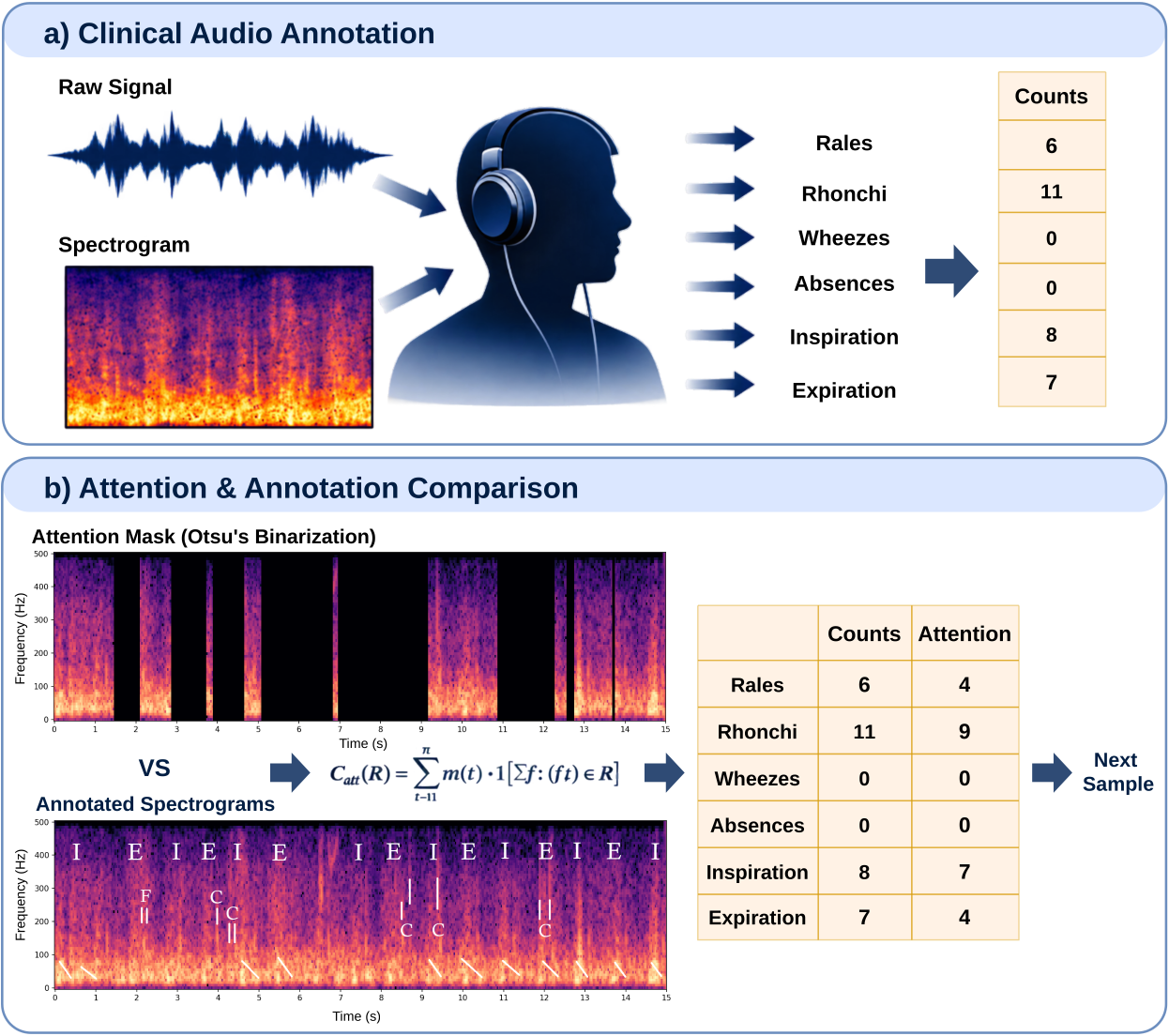}
\caption{
\textbf{a)} Clinical audio annotation workflow. Raw respiratory signals are converted to log-power spectrograms and annotated by experts for adventitious sounds, where diagonal strokes denote rhonchi and vertical strokes denote rales, labeled as \textbf{F} (fine) or \textbf{C} (coarse), while respiratory phases are marked as \textbf{I} = inspiration and \textbf{E} = expiration. 
\textbf{b)} Comparison between expert annotations and model-derived attention maps obtained via Otsu’s binarization. The attention mask highlights spectrotemporal regions of interest, which are quantitatively compared to annotated events. Other marks such as \textbf{A} for absences and circles for wheezes are rare and not present in the illustrated sample.
}

\label{fig:Atten}
\end{figure}

\section{Discussion}

This prospective study shows that automated lung‑sound analysis can discriminate pneumonia from non‑pneumonia in long‑term care residents \cite{Landry2025ERR}. Beyond overall performance, the key contributions are methodological: demonstrating that supervisory label choice strongly shapes model behavior; that a train‑on‑all/infer‑on‑subset channel strategy preserves discrimination while enabling a practical bedside protocol; and that patient‑level partitioning with multi‑seed interpretability yields physiologically coherent and reproducible acoustic evidence. Collectively, these results help close the gap between inflated in‑dataset claims and the more modest but reliable performance seen under rigorous prospective evaluation \cite{cabitza2026almost}.

Models supervised with radiographic (XRAY) labels were more stable and discriminative than those using clinical‑diagnosis (Diag) labels \cite{wang2024impact}. Clinical diagnosis in LTC is noisy: atypical presentations (delirium, functional decline) often replace classic respiratory signs, and nursing‑home assessments show substantial inconsistency \cite{Mylotte2020JAMDA}. Overdiagnosis is common; for example, a multicenter study found that 12\% of adults treated for community‑acquired pneumonia lacked diagnostic justification \cite{gupta2024inappropriate}. Thus, Diag labels conflate true pneumonia with mimics and heuristics, introducing label noise that harms generalization \cite{sukhbaatar2014learning}. Although chest radiography is imperfect, showing fair–moderate interobserver agreement and often trailing clinical symptom onset, it provides a more reproducible anchor; modest reductions in label noise improve discrimination and stability \cite{albaum1996interobserver, tang2021data, bruns2010pneumonia}. Consistent with this, the stethoscope‑only model achieved its strongest and least variable performance (F1 = 0.729; accuracy = 0.783) under XRAY supervision. Structured clinical variables mainly sharpened decision thresholds (fusion AUC = 0.791; specificity = 0.867), suggesting acoustics drive discrimination while clinical data refine operating points.

Channel ablations suggest a practical remedy for long‑term care (LTC) constraints: although comprehensive multi‑site protocols are often infeasible for bedridden or cognitively impaired residents because posterior sites require repositioning and prolonged contact \cite{heitmann2023deepbreath, Landry2025ERR}, models trained on all six channels can be deployed using only the three or four most informative anterior sites while preserving—or slightly improving—performance (Test 3/4: F1 0.736/0.733; specificity 0.895/0.863) and reducing inter‑split variability. Consistent with prior multi‑channel work showing increasing discriminative value from spatial coverage \cite{kim2025enhanced}, our work indicates that broad spatial coverage during training improves learning even when deployment uses fewer sites \cite{messner2020multi}. Interpretability analyses reinforce an anterior‑site protocol with multi‑seed Borda aggregation ranked channels 3 and 4 highest. Moreover, these mid‑thoracic anterior sites span different lobes and can capture broader physiologic information \cite{standring2005gray}. Finally, posterior auscultation remains clinically preferred, yet in LTC many maneuvers are often infeasible; thus, these anterior sites are quicker, more accessible for supine residents, and likely offer gains from improved accessibility and acoustic stability rather than posture equivalence \cite{fiz2008effect}.

We downsampled to \(f_s = 1000\,\text{Hz}\) (\(f_{\text{Nyquist}} = 500\,\text{Hz}\)) to fully capture coarse crackles and rhonchi, which concentrate below \(\sim 300\,\text{Hz}\), while discarding many high‑frequency artifacts similar to fine crackles; this preserves morphology, improves signal‑to‑noise for diagnostically relevant low‑frequency pathology, and reduces computational cost \cite{reichert2008analysis, ye2022regularity, jin2024study}. Models trained on these signals consistently attended to rhonchi as the primary cue and to crackles as secondary cues, explainable by crackles’ brief transients similar to mishandling and clipping \cite{vyshedskiy2012crackle, bohadana2014fundamentals}. Inspiratory crackles are epidemiologically strongly associated with pneumonia (reported \(\sim 81\%\) in cases vs.\ \(\sim 28\%\) in controls) but suffer from poor inter‑observer reliability and low pooled sensitivity in human auscultation. Thus, automated analysis that emphasizes stable, low‑frequency features can improve sensitivity while avoiding over‑reliance on noisy human labels \cite{murphy2004automated, Arts2020SciRep}. (See Extended Data: Domain Shift Analysis for downstream effects on performance.)

Reported accuracies of 90–100\% on public repositories (for example, ICBHI) are difficult to interpret because disease labels often correlate with collection site, equipment, age, or protocol, enabling models to learn acquisition artifacts rather than pathology \cite{cabitza2026almost}. Empirical analyses across multiple datasets have shown that such shortcut learning can inflate apparent performance by roughly 20\%, and correcting for these biases can reduce reported accuracy by up to ~30\% \cite{ong2024shortcut, sukhbaatar2014learning}. Our prospective single‑cohort design uses strict patient‑level partitioning to prevent identity leakage, while multi‑seed evaluation mitigates overfitting, yielding an AUC of approximately 0.77–0.79—substantially lower than public‑dataset claims but closely aligned with other prospective studies such as DeepBreath (internal AUROC 0.75, external 0.74) \cite{heitmann2023deepbreath}.

This study has several limitations. The cohort is modest (185 residents, 73 pneumonia cases), and only the dementia subgroup (N = 45; 14 positives) had sufficient representation for interpretable subgroup analysis; other subgroups (for example, asthma/COPD and tracheostomy) are too small for reliable estimates (see Extended Data Table 3) \cite{gupta2024inappropriate, janssens2004pneumonia}. Chest radiography, while more objective than syndromic diagnosis, is an imperfect reference in frail geriatric patients and contributes unavoidable label noise that likely constrains attainable discrimination \cite{linsalata2020pneumonia, ticinesi2016lung}. A mid‑study DSP change introduced a domain shift between \texttt{before} and \texttt{after} recordings that affected calibration and per‑domain discrimination; the Extended Data: Domain Shift Analysis shows how excluding the \texttt{before} domain, adversarial adaptation, and sampling‑rate sensitivity influence robustness. Finally, this single‑center study enrolled only patients with respiratory symptoms plus fever and lacked external validation, so the proposed anterior three/four‑channel protocol requires confirmation in independent, multi‑center, cross-population cohorts before its feasibility and accuracy can be considered established \cite{park2023methods, park2021key}.

Future work should include externally validated, multi‑site studies in independent nursing‑home cohorts that are adequately powered for clinically important subgroups such as dementia and reported under contemporary AI standards (for example, TRIPOD‑AI and STARD‑AI) \cite{sounderajah2025stard, de2025adherence}. Prospective, comparative evaluations against bedside reference standards—particularly lung ultrasound rather than radiography alone—would clarify the clinical role of automated auscultation in resource‑limited long‑term care pathways \cite{linsalata2020pneumonia}. Using higher‑quality composite references (for example, adjudicated radiography or ultrasound‑supported labels) may raise the attainable performance ceiling \cite{wang2024impact}. Demonstrating clinical utility requires prospective, outcome‑oriented trials that move beyond discrimination metrics to test whether model‑assisted assessment reduces unnecessary transfers, antibiotic overuse, or diagnostic delay \cite{gupta2024inappropriate}, while simultaneously advancing through the SaMD regulatory pathways needed for clinical deployment \cite{muehlematter2021approval}.

\section{Methods}

\subsection{Dataset Acquisition}
\label{DA}

Our study collected lung sounds from patients with suspected pneumonia presenting with upper respiratory symptoms from December 2025 to June 2026, conducted under a protocol approved by the University of Tsukuba IRB (approval no. 2151) and not prospectively registered.

\paragraph{Participants.} Participants were adults $\geq 80$ years receiving care at a skilled nursing home in Ibaraki Prefecture, Japan. Clinical suspicion was based on cough, sputum, or nasal discharge accompanied by fever \(>37.5^\circ\)C. Patients with cognitive impairment were excluded unless a legally authorized representative could consent. All participants provided informed consent either directly or through a legally authorized representative and were not involved in study design.

\paragraph{Clinical Reference and Labels.} Chest radiographs were reviewed by a radiologist for pneumonia (XRAY label), and clinicians assigned clinical diagnoses per Japanese Respiratory Society guidelines (Diag label) \cite{MUKAE2025811}. Labels integrated vitals, auscultation, laboratory/microbiology, differential assessment, and radiographic interpretation. XRAY labels reflected radiographic evidence observe by staff radiologists; Diag labels represented full clinical adjudication.

\paragraph{Digital Stethoscope.} Trained nurses recorded multi‑site anterior lung sounds using the Eko Core 2 digital stethoscope. Nurses completed standardized training emphasizing stable skin contact, minimal movement, avoidance of clothing interference, and quiet environments. Recordings were obtained through the Eko mobile application in clinical mode at a fixed 4\,kHz export rate, retaining native front-end filtering and gain control. Formal acoustic characterization of the Core~2 is limited because microphone sensitivity, noise floor, frequency-response curves, and DSP configuration are proprietary. A subset ($N=45$) captured via on‑device screen recording bypassed parts of the device pipeline; these recordings were retained. No algorithmic noise suppression or post-processing was applied beyond the device’s native conditioning.

\paragraph{Auscultation Protocol} As shown in Figure~\ref{fig:mp} section (a), six anterior chest locations were auscultated in a fixed order: Top Right~(1), Top Left~(2), Middle Right~(3), Middle Left~(4), Bottom Right~(5), and Bottom Left~(6). After protocol standardization, each site was recorded for 15 s. Early 30 s recordings were clipped to remove edge artifacts, typically yielding $\approx28s$ of usable signal. Participants were asked to sit upright and follow a standardized breathing pattern; when not possible, recordings were taken in semi‑Fowler, supine, or lateral positions and documented.

\paragraph{Metadata and Clinical Variables.} Vital signs (blood pressure, temperature, pulse, respiratory rate, SpO\(_2\)) and demographic and clinical metadata (sex, age, body weight/build, respiratory symptoms, pain severity, antibiotic use, comorbidities) were recorded. Acquisition logs documented recording site, duration, patient position, and protocol deviations.

\paragraph{Sample Size Justification} We used all eligible acoustic data, yielding a cohort of 185 geriatric patients (73 pneumonia, 112 non‑pneumonia; 77 with radiographic opacities). This sample supported robust internal model evaluation with the positive pneumonia sample being comparable to or larger than many prior diagnostic‑audio cohorts, especially in LTC settings. With \(n_{+}=73\), the binomial sensitivity SE gives a 95\% CI half‑width of \(1.96\sqrt{S(1-S)/73}\), roughly 7–9 percentage points. Specificity precision with \(n_{-}=112\) produces a slightly tighter 6–8‑point interval. Given these margins and our study objective, the sample size was sufficient without a formal prospective power calculation.

\subsection{Signal Processing and Feature Extraction}
\label{SP}

Following Fig.~\ref{fig:mp}a, all recordings were resampled from thier captured sampling rate to a common $f_s=1000$\,Hz using \texttt{librosa}'s default \texttt{resampy} backend with a zero‑phase Kaiser‑windowed sinc anti‑aliasing filter and polyphase resampling. Waveforms were de‑meaned (DC offset removed) and variance‑normalized per recording (zero mean, unit variance). Each 15\,s and 28\,s recording ($\approx 15{,}000$ samples, $\approx 28{,}000$ samples) was segmented into fixed-length windows of $W=8192$ samples with hop size $H=4096$ (50\% overlap), with the final window shifted so that its endpoint aligns exactly with the signal endpoint—introducing additional terminal overlap—and yielding three or five overlapping windows per recording; thus, being robust to recording length without introducing artifact.

Each window was transformed using a Short-Time Fourier Transform (STFT) with frame length $L = 128$ and hop $b = 32$, producing a frequency resolution of $\Delta f = 7.8125$\,Hz and $N = 256$ time frames. Only the first 64 frequency bins (0--500\,Hz) were retained due to Hermitian symmetry, and each window produced a min–max normalized $64\times256$ log-power spectrogram, where the time axis was fixed to 256 frames to ensure uniform model input dimensionality. Further implementation details (resampling method, padding/truncation rules, STFT/FFT parameters, and normalization formulas) are provided in the Extended Data.

\subsection{Model Development}
\label{modeldev}

All models used the ARP–N convolutional backbone, previously validated on biological acoustics (e.g., blue‑whale D‑call detection). ARP–N operates on log‑power spectrograms and was chosen due to CRNN and CANN variants yielding marginal or negative gains under comparable conditions \cite{rasmussen2025ecologically}.

Patient‑level splits were used throughout to prevent data leakage, with model selection confined to the inner loop. We addressed class imbalance with patient‑level weights: for $N_{+}$ pneumonia and $N_{-}$ non‑pneumonia patients ($N = N_{+} + N_{-}$), we set $w_{+} = N/(2N_{+})$ and $w_{-} = N/(2N_{-})$. These weights were applied to each window’s loss, with each window inheriting its patient’s class weight (no resampling).

For each channel $c$, we restricted inference to that channel and computed the patient‑level probability as \[ \hat{p}^{\,c}_{\text{patient}} = \frac{1}{W_c}\sum_{j=1}^{W_c} \hat{p}^{\,c}_{j}, \] where $W_c$ is the number of windows from channel $c$ and $\hat{p}^{\,c}_{j}$ the window‑level probability. Once cross vlaidation is complete, we averaged fold accuracies $a_{c,F}$ over $N_F$ folds to obtain $\bar{a}_c = \frac{1}{N_F}\sum_{F=1}^{N_F} a_{c,F}$ (here $N_F = 5$), and ranked channels to produce $C^{*}$.

Using the top‑K channels and selecting all windows per channel, each patient contributed up to \(\approx 3K\) or \(5K\) windows (e.g., \(K=6 \rightarrow \approx 18\) or \(30\) windows) with each time–frequency window treated independently (Fig. \ref{fig:mp}c). We aggregated window probabilities as \[ \hat{p}_{\text{patient}} = \frac{1}{N}\sum_{i=1}^{C}\sum_{j=1}^{W_i} \hat{p}_{i,j},\] where $N=\sum_{i} W_i$ is the total number of windows per patient. The model's native decision threshold was used at $.5$. Because training always used all channels, outer loop test‑time masking simply omitted non‑selected channels without altering the input structure \cite{rasmussen2026channel}.

\paragraph{Borda Scoring}

We derived a stable, physiologically interpretable channel hierarchy by applying Borda aggregation across the $5 \times 5$-fold patient‑level cross‑validations. Each of the 25 folds produced an independent ranking of the six channels. For a given fold $f \in \{1,\dots,25\}$, let $r_f(c)$ denote the rank of channel $c$ (1 = best, 6 = worst). We converted ranks to Borda scores via $b_f(c) = 7 - r_f(c)$, so ranks 1–6 map to scores 6–1. The final importance score for each channel was obtained by summing its Borda points across all folds:

\[
B(c) = \sum_{f=1}^{25} b_f(c).
\]

This yields a cross‑validated, seed‑averaged channel importance that reduces fold‑specific variability and prevents single‑iteration ‘lucky’ rankings from dominating. The resulting $B(c)$ values provide an interpretability-focused channel hierarchy that are not used for outer-loop channel masking.

\paragraph{Channel and MIL Architectures.}
Channel‑aware baselines were: (1) a channel-stacked model concatenating six sites into a multi‑channel tensor with spatial filters, and (2) Multi Instance Learning that treats each window as an independent instance with neutral network based patient‑level aggregation  \cite{messner2020multi, nguyen2026lung}. These architectures were evaluated against the ARP--N mean-probability baseline.

\paragraph{Multimodal Extension.}
To incorporate structured clinical information, ARP--N was extended with a parallel clinical branch. We included structured variables (cough, sputum, dyspnea, $\mathrm{SpO}_2$, respiratory rate, fever \cite{vaughn2024community}) and handled a single missing $\mathrm{SpO}_2$ by imputing the cohort mean. The acoustic pathway produced a 128‑D embedding; clinical variables passed through a two‑layer MLP to yield a matching 128‑D embedding.

\paragraph{Hyperparameters} All models were trained under a fixed set of hyperparameters, with the only varying component being the channel‑selection procedure imposed by the inner loop. The inner loop constrained the subset of channels available to each outer‑fold training run, while all other optimization settings—including batch size, learning rate schedule, regularization, and early‑stopping criteria—were held constant across experiments. Importantly, all models were optimized for the best F1 score attainable.

\subsection{Annotation and Attention Map Construction}
\label{annotation}

As illustrated in Figure~\ref{fig:Atten}, recordings were converted into STFT representations prior to annotation. Four annotators independently reviewed each sample: two nurses (A,B), a signal‑domain expert (C), and a pulmonary nurse practitioner with decades of experience (D). They used both the resampled audio and the time–frequency spectrogram (Fig. \ref{fig:mp}a). Annotators were blinded to all clinical information (diagnosis, labels, demographics, model outputs) and were provided only with patient and channel identifiers to prevent bias. Annotators marked acoustic events on the spectrogram (and referenced audio) labeling inspirations, expirations, rales/crackles (fine/coarse), rhonchi, wheezes, and absence events (atypical silence or missing expected breath sounds). Annotators were not required to mark precise onset/offset times; instead they delineated time–frequency regions of perceptual events (spatial regions on the spectrogram). Multiple rales with-in 1 second of each other were considered a single event. Free-text notes were permitted to document noise, motion artifacts, clipping, or other signal degradation.

Model attention was computed using Grad-CAM on the model's final layer for each spectrogram window \cite{selvaraju2017grad}, resized to the original dimensions, and stitched across overlapping windows by averaging in shared regions. A temporal attention curve was obtained by collapsing the stitched map along frequency by taking the maximum, and Otsu's thresholding produced a binary temporal mask \cite{otsu1979threshold}. Annotated events (time–frequency regions) were considered attended if any temporal column in the binary mask overlapped the region; partial overlaps counted as attended. We computed attended counts as 
\[ C_{\mathrm{att}}(R)=\sum_{t=1}^{T} m(t)\,\mathbf{1}\!\left[\exists f : (f,t)\in R\right],\] 
where $m(t)$ is the binary temporal mask and $T$ the number of time columns.

Because all four annotators labeled the same patient/channel combinations, attention alignment was computed separately for each annotator and then aggregated. For every feature type, we devided the model’s attended time columns $C_{\text{att}}(R)$ (Section~\ref{annotation}) by the annotator-provided event counts, yielding per-annotator attention ratios for inspirations, expirations, rales/crackles, rhonchi, wheezes, and absence events. This allowed attention counts to be evaluated both with respect to individual annotators and with respect to shared perceptual regions across annotators, providing a unified measure of how closely the model’s attended time indices aligned with human-identified acoustic events.

\subsection{Statistical Analysis and Reporting Standards}
\label{Stats}

\paragraph{Cohort's Descriptive Analysis} Cohort characteristics were compared between pneumonia and non‑pneumonia groups using standard descriptive analysis. Continuous variables were summarized using means, standard deviations, medians, and interquartile ranges. Categorical variables were summarized using counts and proportions. To estimate p‑values for univariate associations with pneumonia, we applied a robust two-tailed logistic regression model to each predictor, both continuous and categorical, using numeric coercion and dummy encoding; the model was used solely to obtain significance estimates rather than to generate predictive performance.

\paragraph{Model Evaluation} Model performance was estimated using repeated patient‑level cross‑validation following a fixed \(5 \times 5\) design, yielding 25 independent test evaluations. From patient‑level probabilities we computed fold‑level metrics, sensitivity‑constrained operating points (S90 and S80 denote specificity while achieving 90\% and 80\% sensitivity on validation), and calibrated diagnostic quantities (PPV, NPV). AUC was computed with DeLong’s method. Formal mathematical definitions of all evaluation metrics, cross‑validation variability calculations, and diagnostic quantities are provided in the Extended Data. The study adhered to TRIPOD‑AI \cite{collins2024tripod}, including transparent reporting of model architecture, data partitioning, outcome definitions, and uncertainty estimates. External validation and powered subgroup analyses were not feasible; we mitigated this limitation by reporting repeated cross‑validation results for robust internal evaluation.

\section*{Data Availability}

Raw lung‑sound recordings, clinical metadata, and protocols were collected under IRB approval and cannot be publicly released because raw audio and clinical data pose privacy and re‑identification risks. A minimal, fully de‑identified derivative dataset will be available under controlled access and will include log‑power spectrograms, the structured clinical variables used in the multimodal model, and preprocessing/STFT documentation. Editors and reviewers may request access and will receive it within 10 business days after signing a data‑use agreement prohibiting re‑identification, redistribution, and secondary use. Upon acceptance, the dataset will be deposited in a controlled‑access repository for non‑commercial research for at least five years, subject to a brief proposal and agreement to ethical and legal restrictions. No third‑party proprietary datasets were used.

\section*{Code Availability}

All custom code for preprocessing, spectrogram generation, model development (ARP–N, multimodal fusion, channel selection), cross‑validation, and attention‑map construction will be provided to editors and reviewers on request during peer review. Upon acceptance, the complete codebase will be published on a dedicated GitHub repository with documentation and versioning to support reproducibility. No proprietary third‑party code beyond standard open‑source libraries was used, and all algorithmic components central to the study’s conclusions will be included in the public release.

\backmatter

\bmhead{Supplementary Information}

Supplemental information includes Extended Data with full methodological details and mathematical formulations, ablations on capture fidelity and sampling‑rate tuning, stability analysis of the top channel‑selection model, and subgroup performance results. Additional materials such as annotations, attention maps, and annotation counts are available on request.

\bmhead{Acknowledgments}

This work was supported in part by Amazon’s Cross‑Pacific Artificial Intelligence Initiative. Funding was provided as a corporate gift, with salary support administered through the University of Washington. We gratefully acknowledge this contribution, which enabled development of the digital‑auscultation pipeline, multimodal modeling framework, and evaluation infrastructure.

AI tools (large‑language models) assisted with structural writing, language refinement, and non‑scientific symbolic figure assets. All scientific images (e.g., spectrograms) are real and unaltered. The tools provided light assistance in summarizing patterns present in the quantitative results; all statistical analyses, methodological decisions, and scientific interpretations were performed and verified by the authors. No AI system contributed to study design, data collection, model development, or final decision‑making on the manuscript’s conclusions.

We also wish to acknowledge the late Dr. Les Atlas, who served as the original Co‑Investigator on this project. Before his passing, Dr. Atlas provided foundational insight into key digital signal processing methods and strongly advocated for the support that made this work possible. His intellectual generosity, mentorship, and commitment to advancing acoustic analysis in medicine shaped the trajectory of this research, and we are honored to continue the work he helped initiate.

\section*{Author Contributions}

\noindent \textbf{N.R.} designed the majority of the experimental framework, conducted dataset analysis, implemented all computational and methodological advancements, and led the writing of the manuscript. He additionally contributed to dataset organization, stethoscope protocol analysis, and performed clinical sound annotation.

\noindent \textbf{O.Z.} provided extensive manuscript review, oversaw the University of Washington clinical team, and contributed substantially to the writing—particularly the discussion and clinical interpretation.

\noindent \textbf{Z.L.W.} assisted in dataset collection and organization, manuscript writing, figure development, and annotating clinical sounds.

\noindent \textbf{H.Y.} contributed to dataset collection, manuscript writing, figure development, and annotating clinical sounds.

\noindent \textbf{J.T.F.} conceptualized the clinical interpretation of adventitious sounds, provided critical physiological commentary and review, and annotated clinical sounds.

\noindent \textbf{K.N.} ensured adherence to digital signal processing conventions, contributed to discussion commentary, and reviewed the manuscript.

\noindent \textbf{A.K.} contributed to the data-collection protocol design, supported the Japan data-collection site, and reviewed the manuscript.

\noindent \textbf{T.I.} oversaw data collection at the long-term care facility in Japan and provided methodological feedback.

\section*{Competing Interests}

N.R. previously received salary support from Virufy (Covid Detection Foundation), a non‑profit organization developing AI‑based respiratory diagnostic tools founded by A.K. Although A.K., affiliated with Virufy, contributed to aspects of the study design and manuscript review in his capacity as a co‑author, Virufy as an organization provided no funding and had no role in directing the study design, data collection, analysis, interpretation, or manuscript preparation. All other authors declare no competing financial or non‑financial interests.

\clearpage

\begin{table}[t!]
\tiny
\centering
\begin{tabularx}{\textwidth}{|>{\columncolor{gray!15}}l|lX|lX|Xl|}
\multicolumn{7}{c}{\textbf{Continuous Variable Summary for Non-Pneumonia and Pneumonia with Hypothesis Testing}} \\ 
\hline
\multicolumn{1}{|c}{} & \multicolumn{2}{|c|}{\textbf{Non-Pneumonia}} & \multicolumn{2}{c|}{\textbf{Pneumonia}} & \multicolumn{2}{c|}{\textbf{Hypothesis Testing}} \\
\hline
\textbf{Variable} &
\textbf{Mn \(\pm\) SD} & \textbf{Med (IQR)} & \textbf{Mn \(\pm\) SD} & \textbf{Med (IQR)} & \textbf{OR\(|\)CI} & \textbf{P} \\
\hline
Age (Years) & \(86.1 \pm 10.4\) & 88.8 (8.98) & \(84.4 \pm 10.6\) & 85.9 (12.2) & .984\textbar .957,1.01 & .271 \\
Weight (Kg) & \(45.3 \pm 10.7\) & 44 (10.4) & \(47.1 \pm 12.3\) & 45 (10.9) & 1.01\textbar .978,1.05 & .444 \\
Oxygen Saturation (\%) & \(95.7 \pm 2.76\) & 96 (3) & \(94.7 \pm 3.67\) & 95 (4) & .907\textbar .822,1 & .05 \\
Pulse (Beats/Min) & \(76.6 \pm 13.7\) & 77 (21.2) & \(81.7 \pm 15.5\) & 79 (21) & 1.03\textbar 1.00,1.04 & .02 \\
Systolic  BP (mmHG) & \(118. \pm 19.5\) & 120 (29.5) & \(121. \pm 21.7\) & 121 (29) & 1.01\textbar .992,1.02 & .381 \\
Diastolic  BP (mmHG) & \(69.5 \pm 13.9\) & 69 (21) & \(73.1 \pm 10.9\) & 74 (17) & 1.02\textbar .998,1.04 & .068 \\
Resp. Rate (Breaths/Min) & \(15.9 \pm 4.96\) & 14 (8) & \(16.6 \pm 4.9\) & 16 (6) & 1.03\textbar .968,1.09 & .37 \\
Body Temp (Celcius) & \(36.4 \pm .46\) & 36.4 (.7) & \(36.5 \pm .64\) & 36.5 (.9) & 1.44\textbar .823,2.50 & .203 \\
Visual Analog Scale (0-100) & \(22.8 \pm 18.4\) & 20 (12.5) & \(34.8 \pm 26.7\) & 30 (30) & 1.02\textbar 1.01,1.03 & .001 \\
\hline
\end{tabularx}

\vspace{6pt}

\begin{tabularx}{\textwidth}{|>{\columncolor{gray!15}}lX|XX|XX|XX|}
\multicolumn{8}{c}{\textbf{Categorical Variable Summary for Non-Pneumonia and Pneumonia Diagnoses with Hypothesis Testing}} \\ 
\hline
\multicolumn{2}{|c|}{} & \multicolumn{2}{c|}{\textbf{Non-Pneumonia}} & \multicolumn{2}{c|}{\textbf{Pneumonia}} & \multicolumn{2}{c|}{\textbf{Hypothesis Testing}} \\
\hline
\textbf{Variable} & \textbf{Category} & \textbf{N} & \textbf{\%} & \textbf{N} & \textbf{\%} & \textbf{OR\textbar CI} & \hspace*{6pt} \textbf{P} \\
\hline

Body Type & Normal \newline Underweight \newline Overweight & 28 \newline 81 \newline 3 & 25 \newline 72 \newline 3 & 20 \newline 49 \newline 4 & 27 \newline 67 \newline 5 & - \newline .847\textbar .431,1.66 \newline 1.87\textbar .376,9.27 & \hspace*{6pt} - \newline \hspace*{6pt} .629 \newline \hspace*{6pt} .445 \\
Gender & Female \newline Male & 55 \newline 57 & 49 \newline 51 & 34 \newline 39 & 47 \newline 53 & - \newline 1.11\textbar .613,1.99  & \hspace*{6pt} - \newline \hspace*{6pt} .736 \\
Cough Presence & Few \newline Often & 67 \newline 14 & 60 \newline 12 & 44 \newline 7 & 60 \newline 10 & .925\textbar .476,1.80 \newline .705\textbar .244,2.03 & \hspace*{6pt} .819 \newline \hspace*{6pt} .517 \\
Dyspnea & Activity \newline Constant & 75 \newline 11 & 67 \newline 10 & 39 \newline 19 & 53 \newline 26 & .901\textbar .428,1.89 \newline 2.99\textbar 1.12,7.95 & \hspace*{6pt} .784 \newline \hspace*{6pt} .028 \\
Nasal Color & Clear \newline Yellow & 3 \newline 1 & 3 \newline 1 & 0 \newline 0 & 0 \newline 0 & 0\textbar 0,inf \newline 0\textbar 0,inf & \hspace*{6pt} .999 \newline \hspace*{6pt} 1 \\
Nasal Consistency & Thin \newline Thick & 2 \newline 2 & 2 \newline 2 & 0 \newline 0 & 0 \newline 0 & 0\textbar 0,inf \newline 0\textbar 0,inf & \hspace*{6pt} 1 \newline \hspace*{6pt} 1 \\
Sputum Color & Clear \newline White \newline Yellow \newline Bloody & 4 \newline 30 \newline 24 \newline 1 & 4 \newline 27 \newline 21 \newline 1 & 2 \newline 31 \newline 21 \newline 3 & 3 \newline 42 \newline 29 \newline 4 & 1.66\textbar .277,9.89 \newline 3.42\textbar 1.61,7.25 \newline 2.90\textbar 1.29,6.51 \newline 9.94\textbar .966,inf & \hspace*{6pt} .58 \newline \hspace*{6pt} .001 \newline \hspace*{6pt} .01 \newline \hspace*{6pt} .054 \\
Sputum Consistency & Thin \newline Thick & 7 \newline 52 & 6 \newline 46 & 4 \newline 53 & 5 \newline 73 & 1.89\textbar .491,7.29 \newline 3.38\textbar 1.71,6.64 & \hspace*{6pt} .354 \newline \hspace*{6pt} .001 \\
Tracheostomy & True & 10 & 9 & 1 & 1 & .142\textbar .018,1.13 & \hspace*{6pt} .065 \\
CO Asthma & True & 9 & 8 & 5 & 7 & .842\textbar .27,2.61 & \hspace*{6pt} .766 \\
CO COPD & True & 2 & 2 & 5 & 7 & 4.04\textbar .763,21.4 & \hspace*{6pt} .101 \\
CO Cancer & True & 12 & 11 & 12 & 16 & 1.64\textbar .693,3.87 & \hspace*{6pt} .261 \\
CO Heart Failure & True & 7 & 6 & 4 & 5 & .870\textbar .245,3.08 & \hspace*{6pt} .829 \\
CO Int. Pneumonia & True & 0 & 0 & 2 & 3 & inf\textbar 0,inf & \hspace*{6pt} 1 \\
CO Dementia & True & 36 & 32 & 23 & 32 & .971\textbar .515,1.82 & \hspace*{6pt} .928 \\

\hline
\end{tabularx}

\caption{Continuous and categorical clinical characteristics of the study cohort, stratified by non‑pneumonia and pneumonia diagnoses, reference categories such as `None' or `False' are not included. Continuous variables are reported as mean ± standard deviation and median (interquartile range). Categorical variables are reported as counts and percentages. Two-tailed univariate logistic regression provides odds ratios with 95\% confidence intervals and associated p‑values for each variable’s association with pneumonia where omission denotes reference variables.}
\label{Table:Cohort}
\end{table}

\clearpage

\begin{table*}[t]
\centering
\tiny
\begin{tabularx}{\textwidth}{|>{\columncolor{gray!25}}lXXXXXXXXXXXXXXXXXX}
\toprule
\textbf{Model}
& \textbf{F1}
& \textbf{Acc}
& \textbf{AUC}
& \textbf{Spec}
& \textbf{NPV}
& \textbf{Sens}
& \textbf{PPV}
& \textbf{S90}
& \textbf{S80} \\
\midrule

% ===================== DIAG LABEL =====================
\multicolumn{10}{>{\columncolor{gray!35}}l}{\textbf{Diag Label — Clinical Variables}} \\

Avg
& .585 & .688 & .666 & .771 & .730 & .559 & .641 & .223 & .370 \\

Inter StDev
& .007 & .021 & .021 & .051 & .004 & .027 & .049 & \textbf{.036} & .052 \\

Intra StDev
& .101 & .088 & \textbf{.087} & .128 & .071 & \textbf{.131} & .155 & \textbf{.120} & \textbf{.170} \\

\multicolumn{10}{>{\columncolor{gray!35}}l}{\textbf{Diag Label — Stethoscope Only}} \\

Avg
& \textbf{.637} & .711 & .719 & .756 & \textbf{.771} & \textbf{.639} & .683 & .267 & .401 \\

Inter StDev
& \textbf{.011} & .022 & \textbf{.010} & .067 & .021 & .049 & \textbf{.038} & .069 & \textbf{.034} \\

Intra StDev
& \textbf{.086} & .090 & .099 & .192 & \textbf{.066} & .145 & .175 & .190 & .206 \\

\multicolumn{10}{>{\columncolor{gray!35}}l}{\textbf{Diag Label — Clinical + Stethoscope}} \\

Avg
& .632 & \textbf{.726} & \textbf{.730} & \textbf{.809} & .759 & .601 & \textbf{.687} & \textbf{.288} & \textbf{.462} \\

Inter StDev
& .023 & \textbf{.017} & .017 & \textbf{.028} & \textbf{.011} & \textbf{.023} & .040 & .073 & .043 \\

Intra StDev
& .106 & \textbf{.074} & .096 & \textbf{.110} & \textbf{.067} & .133 & \textbf{.124} & .199 & .236 \\

\midrule
% ===================== XRAY LABEL =====================
\multicolumn{10}{>{\columncolor{gray!35}}l}{\textbf{XRAY Label — Clinical Variables}} \\

Avg
& .644 & .707 & .689 & .763 & .741 & .629 & .670 & .236 & .488 \\

Inter StDev
& .027 & .029 & .019 & .047 & .015 & \textbf{.019} & .047 & .030 & .058 \\

Intra StDev
& .086 & .082 & .105 & \textbf{.116} & .056 & .081 & .124 & .239 & .237 \\

\multicolumn{10}{>{\columncolor{gray!35}}l}{\textbf{XRAY Label — Stethoscope Only}} \\

Avg
& \textbf{.729} & \textbf{.783} & .774 & .851 & \textbf{.789} & \textbf{.686} & \textbf{.805} & .262 & .580 \\

Inter StDev
& \textbf{.017} & .026 & \textbf{.016} & .085 & .015 & .065 & .055 & \textbf{.059} & .051 \\

Intra StDev
& \textbf{.067} & \textbf{.071} & .083 & .129 & \textbf{.050} & \textbf{.076} & \textbf{.116} & \textbf{.188} & .247 \\

\multicolumn{10}{>{\columncolor{gray!35}}l}{\textbf{XRAY Label — Clinical + Stethoscope}} \\

Avg
& .685 & .768 & \textbf{.791} & \textbf{.867} & .773 & .629 & .803 & \textbf{.375} & \textbf{.608} \\

Inter StDev
& \textbf{.017} & \textbf{.013} & \textbf{.016} & \textbf{.017} & \textbf{.013} & \textbf{.019} & \textbf{.024} & .112 & \textbf{.046} \\

Intra StDev
& .116 & .073 & \textbf{.078} & .118 & .066 & .160 & .132 & .200 & \textbf{.209} \\

\bottomrule
\end{tabularx}
\caption{Main model results across Diag and XRAY labels for Clinical, Stethoscope, and Fusion models. Values are averaged across five patient-level splits. Avg = mean of Cross-Validation means; Inter StDev = standard deviation of Cross-Validation means; Intra StDev = mean of per-Cross-Validation standard deviations.}
\label{tab:MR}
\end{table*}

\clearpage


%\bibliographystyle{unsrt}
\bibliography{sn-bibliography}% common bib file

\end{document}